\documentclass[conference]{IEEEtran}
\IEEEoverridecommandlockouts
\usepackage[letterpaper, left=0.75in, right=0.75in, bottom=0.75in, top=1in]{geometry}
\usepackage{cite}
\usepackage{multicol, blindtext}
\usepackage{amsmath,amssymb,amsfonts}
\usepackage{graphicx, bbm}
\usepackage{textcomp, nccmath, cuted}
\usepackage{algorithm}
\usepackage{algorithmic} 
\usepackage{afterpage}
\usepackage{amsfonts} 
\usepackage{comment}
\usepackage{xcolor}
\def\BibTeX{{\rm B\kern-.05em{\sc i\kern-.025em b}\kern-.08em
    T\kern-.1667em\lower.7ex\hbox{E}\kern-.125emX}}
    
\begin{document}

\newgeometry{ left=0.75in, right=0.75in, bottom=0.75in, top=0.75in}
\addtolength{\topmargin}{0.1in}
\title{On the Importance of Trust in Next-Generation Networked CPS Systems: An AI Perspective}

\author{\IEEEauthorblockN{Anousheh Gholami$^*$, Nariman Torkzaban$^*$, and John S. Baras
\thanks{$^*$ The identified authors contributed equally to this paper. Their names appear according to alphabetical order.}}
\IEEEauthorblockA{\textit{Department of Electrical \& Computer Engineering}\\
\textit{\& Institute for Systems Research}\\
Email: \{anousheh, narimant, baras\}@umd.edu}
}



\maketitle

\begin{abstract}


With the increasing scale, complexity and heterogeneity of the next generation networked systems, seamless control, management and security of such systems becomes increasingly challenging. Many diverse applications have driven interest in networked systems, including large-scale distributed learning, multi-agent optimization, 5G service provisioning, and network slicing, etc. 
In this paper, we propose trust as a measure to evaluate the status of network agents and improve the decision making process. We interpret trust as a relation among entities that participate in various protocols. Trust relations are based on evidence created by the interactions of entities within a protocol and may be a composite of multiple metrics such as availability, reliability, resilience, etc. depending on application context. We first elaborate on the importance of trust as a metric and then present a mathematical framework for trust computation and aggregation within a network. Then we show in practice, how trust can be integrated into network decision making processes by presenting two examples. In the first example, we show how utilizing the trust evidence can improve the performance and the security of Federated Learning. Second, we show how a 5G network resource provisioning framework can be improved when augmented with a trust-aware decision making scheme. We verify the validity of our trust-based approach through simulations. Finally, we explain the challenges associated with aggregating the trust evidence and briefly explain our ideas to tackle them. 

\end{abstract}
\vspace{1mm}

%
\IEEEpeerreviewmaketitle
\section{Introduction}
\label{sec:intro}

\begin{figure*}[ht]
\begin{center}
\begin{minipage}[h]{0.90\textwidth}
\includegraphics[width=1\textwidth]{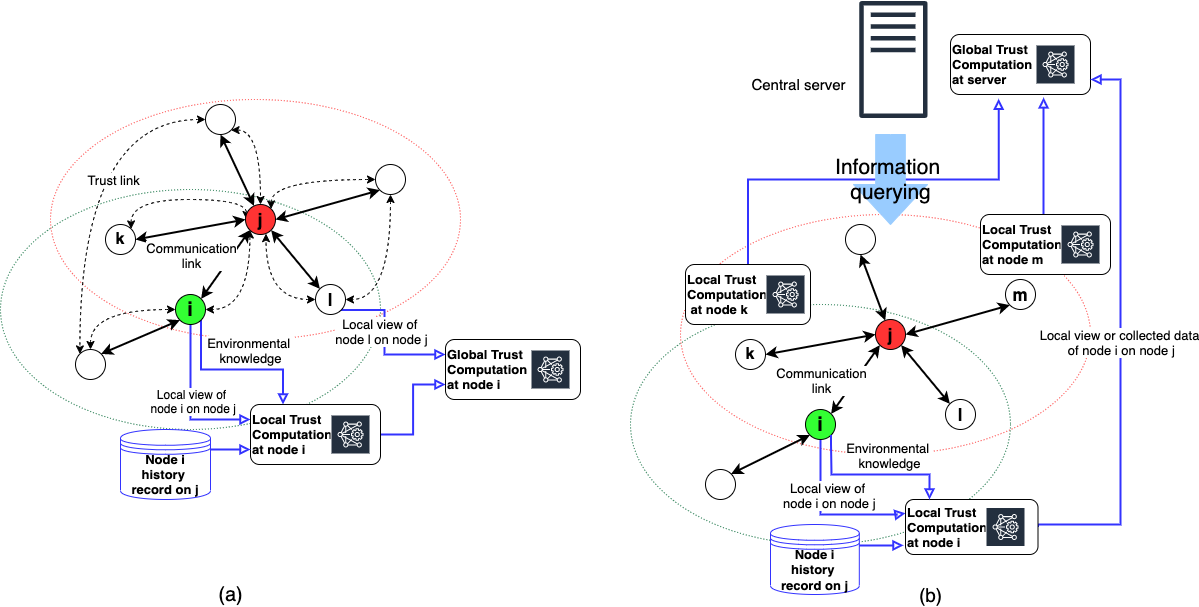}
  \caption{Trust aggregation framework in (a) decentralized and (b) centralized regimes}
  \label{trust}
\end{minipage}
\end{center}
\end{figure*}

Given the increasing complexity of the cyber-physical systems (CPS), the development of a novel framework for modeling, analysing and predicting the behavior of such systems 
is of paramount importance.
With the recent advances in Internet of Things (IoT), and 5G promises to support massive machine-to-machine (M2M) communications, tightly-coupled next-generation CPS devices will be collaborating with more sophisticated sensing, computing and communications capabilities and in much larger scales. Besides the need to a more complicated management and control scheme, with the heterogeneity and the large scale of the CPS systems, dealing with privacy and security threats becomes more central. With more advanced computation and communications capabilities, the collaborative CPS agents are enabled to realize a wide range of applications and use-cases, involving, data collection, processing, and decision making, from healthcare, vehicular networks, and smart manufacturing, to 5G service provisioning, and content delivery. All these applications heavily rely on the constant exchange of collected raw data and the processed information between the collaborating agents, as opposed to the traditional case where data were collected and processed at a centralized entity. The fact that the information is crowd-sourced by the CPS agents, to a large extent, eliminates the risk of the existence of a single point of failure and contributes to the resilience of the network, but at the same time demonstrates the need to establish trust relationships between the agents that are exchanging information. More specifically, apart from ensuring the security of communications between the network agents, it is essential to answer the following questions: (i) whether an agent refuses to share its information with other agents due to privacy concerns or conflict of interest? (ii) whether an agent manipulates the received data before processing? (iii) whether an agent intentionally or unintentionally, shares incorrect information with the rest of the network? etc. \cite{8329991} \cite{1589111}; In other words, it is essential to establish to what extent each agent of the network can be trusted. Clearly such mechanisms of trust contribute essentially in the resilience of networked cyber-physical systems (Net-CPS).

Within the context of Net-CPS, we interpret trust as a relation between different network entities that may interact or collaborate in groups towards achieving various specific goals. These relations are set up and updated based on the evidence generated when the agents collaborate within a previous protocol. If the collaboration has been contributive towards achievement of the specific goal (positive evidence), the parties accumulate their trust perspective towards one another, otherwise (negative evidence), trust will decrease between them. Trust estimates have input to decisions such as access control, resource allocation, agent participation, and so on. The method by which trust is computed and aggregated within the network may depend on the specific application, however similar to \cite{1589111}, we enumerate the central differences in the terminology of how trust computation and aggregation can be approached: 
\begin{itemize}
    \item \textit{Centralized vs. Decentralized}: Under the \textit{centralized} regime, all the network entities rely on a central trusted party that estimates the trustworthiness level of each entity and updates all the network nodes. In this sense, all the nodes are enforced to agree on the degree by which each entity is trusted as dictated by the central provider. On the other hand, under the \textit{decentralized} approach, each user itself is responsible for calculating its opinion on the level of trustworthiness for each entity it might be interested in. This distinction however is irrelevant to the fashion trust is computed and only relates to the semantics of trust. For instance, under a decentralized regime a user may utilize a distributed approach for computing the trust of its target. 
    
    \item \textit{Global vs. Local}: \textit{Local} trust is the opinion that a trustor node has towards a trustee and is generated depending on the first-hand evidence gathered based on local interactions, however, \textit{global} trust  is formed by combining the first-hand evidence and the opinions of other nodes about the specific trustee, and is usually more accurate. In fact the local exchange of the local observations is used towards obtaining the global trust \cite{6916263}.

    \item \textit{Proactive vs. Reactive}: Under a \textit{proactive} regime, the entities manage to keep the trust estimates updates, while under a \textit{reactive} regime, the trust estimates are computed only when they are required. The proactive scheme is not communication efficient as a large bandwidth needs to be consumed to keep the trust values updated; therefore a reactive scheme is usually preferred unless the frequency by which trust decisions are made is comparable to the frequency of the local trust updates.
    
    \item \textit{Direct vs. Indirect:} \textit{Directed} trust is obtained via interaction through direct communication with another agent. However, \textit{indirect} trust is a trust relationship between two entities that have not interacted in the past. Establishing indirect trust relationship, heavily relies on the assumption that trust has the \textit{transitivity} property which is not necessarily the case in any application. 
    
\end{itemize}



\section{Trust Aggregation Model} 
\label{sec:trustAgg}

\subsection{Trust Aggregation Framework}
In this section we present two schemes for propagating and aggregating the trust estimates within a network of CPS devices (\ref{trust}). The first scheme corresponds to the case where there is no central entity involved in estimating the trustworthiness of the network agents, and the nodes participate in \text{direct} computation of trust to obtain \text{local} trust estimates on the other peers, using the locally-available first-hand evidence they have gathered, the record history they have stored from the past observations, and the knowledge they obtain by sensing the environment. Once all agents form their local views, they will participate in local exchanging of their local trust estimates estimates to form the more accurate global values for trustworthiness of the networked agents. Then, the obtained global trust model can be used in the corresponding trust-aware applications. 

Within the second scheme however, the global trust values are obtained in an indirect fashion. There exists a central trusted party that is constantly monitoring the network and is communicating with the CPS agents to gather evidence on their state. The CPS agents may share their local view on their neighbors with the central entity which may be used in computing the trust estimates by the central entity. Once the central party calculates the trust values of the CPS agents using the information it has gathered, it will push the relevant information to each agent. The calculated trust values can be used by the central entity to perform centralized trust-aware decision making, or can be used by each agent to participate in local or distributed trust-aware protocols. 

In what follows, we will formalize the above discussion to mathematically model the processing, propagating and aggregation of the trust values. The components of our model mostly follow and rely on the discussion in \cite{6916263}.

We model the network of agents at time instance $k$, as an undirected graph $\mathcal{G}^{(k)} = (\mathcal{N}^{(k)}, \mathcal{L}^{(k)})$ where  $\mathcal{N}$ is the set of nodes and for $n, m \in \mathcal{N}^{(k)}$, $\mathcal{L}^{(k)}$ contains all links $(m,n)^{(k)}$ where agents $m$, and $n$ can communicate with one another at time instance $k$. We denote this graph as the \textit{communication graph} at time instance $k$. Let $\mathcal{N}_i^{(k)}$ be the set of neighbors of node $i$ at time step $k$. Apart from the communication relationship, we define \textit{local trust} relationships between nodes $i, j \in \mathcal{N}^{{(k)}} $. Let $\tau_{ij}^{(k)}$, and $t_{ij}^{(k)}$ be the local  and global view of node $i$ on trustworthiness of the node $j$ at time instance $k$ in respective order. We may ignore the index $k$ whenever doing so does not lead to confusion.

\subsection{Local Trust Model}

 To formalize the definition of local trust, let us define $X^{(k)}_{ij}$ to be a random variable denoting the reputation that node $j$ has in the perspective of node $i$ in time instance $k$. $X^{(k)}_{ij}$ follows a Beta distribution with parameters $\alpha_{ij}^{(k)}$, and $\beta_{ij}^{(k)}$. Moreover, define $r_{ij}^{(k)} = \alpha_{ij}^{(k)} - 1$, and $s_{ij}^{(k)} = \beta_{ij}^{(k)} - 1$ that determine the number of times up to round $k$, that node $j$'s behavior is benign and malicious in perspective of node $i$, in respective order. The details of how $r_{ij}^{(k)}$, and $s_{ij}^{(k)}$, are obtained depends on the specific scenario and is to be explicitly mentioned in the next sections. We let $\tau_{ij}^{(k)}$ to be precisely the expected value of the reputation random variable in the Beta system $X^{(k)}_{ij}$. Formally, we have: 

\begin{align}
    &f_{X^{(k)}_{ij}}(x; \alpha_{ij}^{(k)},\beta_{ij}^{(k)}) =  (\frac{\Gamma\left(\alpha_{i j}^{(k)}+\beta_{i j}^{(k)}\right)}{\Gamma\left(\alpha_{i j}^{(k)}\right) \Gamma\left(\beta_{i j}^{(k)}\right)}) \cdot \nonumber\\&
    (x^{\alpha_{i j}^{(k)}-1}\left(1-x\right)^{\beta_{i j}^{(k)}-1}) \label{reputation}
\end{align}

\begin{align}
    & \tau_{i j}^{(k)}=\mathbb{E}\left[X_{i j}^{(k)}\right]=\frac{r_{i j}^{(k)}+1}{r_{i j}^{(k)}+s_{i j}^{(k)}+2}\label{expected}
\end{align}
Intuitively, the evolution of $r$, and $s$ parameters needs to be in a way that the more recent information receive more relative importance comparing to the older ones. Therefore, we define $0<\rho_1$ $<$ $\rho_2$ as forgetting factors to control the balance between old and new terms. 

\begin{align}
r_{i j}^{(k+1)} &=\rho_{1} r_{i j}^{(k)}+I_{i j}^{(k+1)} \\
s_{i j}^{(k+1)} &=\rho_{2} s_{i j}^{(k)}+1-I_{i j}^{(k+1)},
\end{align}

The function $I_{i j}^{(k+1)} \in [0, 1]$ models the instantaneous perspective of node $i$ on the behavior of node $j$ in $(k+1)^{th}$ round. 

\subsection{Global Trust Model}

At each instance $k$, within the local trust model, each node $i$ computes its local trust for all nodes $j \in \mathcal{N}_i$ in the communication graph. In order to make more accurate estimates, node $i$ will need to take into account the opinions of other network nodes who have first-hand evidence one node $j$'s behavior. Following the approach in \cite{6916263}, node $i$ computes in an iterative fashion its global trust estimate for node $j$, i.e. $t_{ij}^{(k)}$ using the opinions of its neighbors as: 

\begin{equation}
    t_{i j}^{m}=\left\{\begin{array}{cl}
1 & \text { if } i=j \\
\sum_{l \in \mathcal{N}_{i}, l \neq j} w_{i l} t_{l j}^{m-1} & \text { if } i \neq j
\end{array}\right. \label{global}
\end{equation}

where $w_{il} = \frac{\tau_{il}}{\sum_{l \in \mathcal{N}_{i}, l \neq j} \tau_{i l}}$.

In other words, node $i$ pays more attention to the opinions of those of its neighbors who it trusts more. 
We note again that the global trust computation is an iterative process that is going to be embedded in each iteration of the trust-aware protocol. Therefore, to avoid any confusion we have used the iteration counter $m$ for this process. 
Here, we have dropped the superscript $k$ as we assume the value of local trust remains constant within the loop of computing the global trust. 

In the next two sections, we will show how the above trust framework can be used in practice to enhance the security of the CPS-based protocols. We will select as case studies two recent challenging problems from the state of the art, and consider two corresponding well-known algorithms for them, where security is a big challenge in these protocols. We will then augment those protocols by our trust inference and aggregation framework and show how trust can be used to improve the security of such protocols. 
\section{Case Study I: Trust-aware Federated Learning}
In this section we describe the two considered use cases to show the role and importance of trust in a new breed of intelligent networked systems. In the next generation networked systems with high scale and heterogeneity, it is very impractical for the edge devices (sensors, drones, mobile phones, etc.) to transmit their collected data to a remote data center for centralized machine learning tasks, due to the limited communication resources and privacy constraint. Therefore, the recent trend of high-stake applications such as target recognition in drones, augmented/virtual reality, autonomous vehicles, etc. requires a novel paradigm change calling for distributed, low-latency and reliable ML at the network edge (referred to as edge ML \cite{edge-ml}). Federated Learning (FL) \cite{FED-Google} is a new learning framework that allows multiple agents such as mobile phones, sensors or drones to collaboratively train a common model, e.g. a neural network (NN), without sharing their local data, thereby preserving privacy to a great extent. In the following, we first elaborate on two architectures for FL, namely centralized and decentralized FL and their corresponding challenges, and then discuss the question of evaluating the trust decision $I_{ij}^{(k)}$ introduced in section \ref{sec:trustAgg} for these architectures.
\subsection{Centralized FL}
\begin{figure*}[ht]
\begin{center}
\begin{minipage}[h]{0.65\textwidth}
\includegraphics[width=1\textwidth]{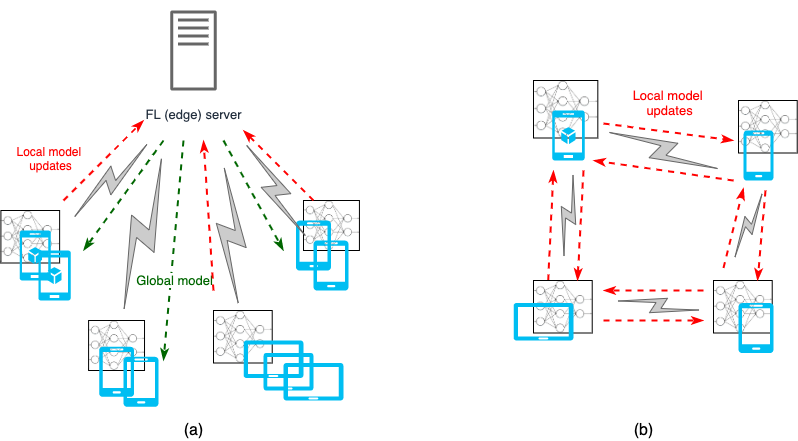}
  \caption{(a) Centralized FL (b) Decentralized FL}
  \label{FL}
\end{minipage}
\end{center}
\end{figure*}
We refer to an FL setting consisting of a central server and multiple users (agents) as centralized FL. In such systems, each agent has a local dataset on which the updates to the current global model are computed, and then the local updates are aggregated and fed-back by a central server. This process is repeated until a desired accuracy level is obtained. Each iteration (communication round) includes a local training phase resulting in the generation of local model parameters and a communication phase to transmit local models to the central server. Indeed, the training process of a typical centralized FL includes the following steps:
\begin{itemize}
    \item \textit{User selection}: The central server selects a set of users at each communication round, considering different factors such as available bandwidth, energy budget, computing power, etc.
    \item \textit{Model broadcast}: The selected users download the most recent global model.
    \item \textit{Local training}: Each selected user performs local computation and generates an update to the model parameter using the local data.
    \item \textit{Model Aggregation}: The central server aggregates the collected local updates and generates a global model.
\end{itemize}
Fig. \ref{FL}(a) illustrates a centralized FL setup. In FL settings, there exist several security issues which mainly arise from the participation of a fleet of possibly unreliable or compromised devices in the training time. According to \cite{FL-tuto}, attacks in FL can be classified into two categories: (i) \textit{data inference attacks}, in which the adversary tries to infer information about the users' private data, and (ii) \textit{model performance attacks}, which includes data poisoning, update poisoning and model evasion attacks. In data poisoning, the adversary subverts the user's dataset in the local training. The goal of the adversary in update poisoning is to alter the model updates transmitted to the central server. The model evasion attack refers to the data alternation at the inference time. In the following, we first describe the federated optimization problem implicit in FL and then explain our proposed trust-based mechanism to defend against the introduced attacks. 

Consider a network consisting of a set $\mathcal{N}$ of $N$ agents and one central server which collaboratively train a learning model. Each agent $i\in \mathcal{N}$ has $D_i$ data samples and the total number of samples is $D$. The $d$th sample is denoted by $(\mathbf{x}_{d},\mathbf{y}_{d})$ where $ \mathbf{x}_{d} \in \mathbb{R}^{D_{in}\times 1}, \mathbf{y}_{d}\in\mathbb{R}^{D_{out}\times 1}$, $d=1,...,D$. We assume that $\mathcal{P}_i$ is the set of indexes of data points on agent $i$ and the data collected by different agents have the same distribution (\textbf{IID} assumption).
The local dataset $ \mathcal{D}_i=\{(\mathbf{x}_d,\mathbf{y}_d), d\in\mathcal{P}_i\}$ is used to train a local model $\mathcal{M}_i$ parameterized by $\mathbf{w}_i$. Let $\mathcal{M}$ and $\mathbf{w}$ be the global model and the global parameter vector. The objective of the FL training is to minimize:
\begin{align}
    F(\mathbf{w}) = \sum_{i \in \mathcal{N}} q_i F_i(\mathbf{w}) = \sum_{i \in \mathcal{N}} q_i \frac{1}{D_i} \sum_{d\in \mathcal{P}_i}  f(\mathbf{w}, \mathbf{x}_{d}, \mathbf{y}_{d}) \label{FLobj}
\end{align}
where $f(\mathbf{w}, \mathbf{x}_{d}, \mathbf{y}_{d})$ is the loss function capturing the accuracy of the FL model, and $q_i$ is the weight of the $i$th device such that $q_i \geq 0$ and $\sum_{i} q_i = 1$. The objective of FL is to minimize \eqref{FLobj}. As a leading algorithm in this setting, \textit{Federated
Averaging (FedAvg)} \cite{mcmahan2017communication} runs Stochastic Gradient Descent (SGD) in parallel on a small subset of the total agents and averages the local updates only once in a while. Although \textit{FedAvg} is showed to stabilize the convergence and ensure the privacy to a great extent, since the central server has limited control over the behaviour of the participating agents, it is vulnerable to adversary behaviours such as poisoning attacks as shown in \cite{xie2019slsgd}. To address this issue, we explore the incorporation of trust into \textit{FedAvg}. The proposed algorithm is referred to as \textit{trusted FedAvg} and is described in the pseudocode of Algorithm \ref{trustedfedavg}.

Let $t_{Si}^{(k)}$ be the opinion of the central server on trustworthiness of agent $i$ at round $k$. The server obtains the value of $t_{Si}^{(k)}$ based on the model described in section \ref{sec:trustAgg}. Moreover, the trust evaluation method proposed for FL setups is discussed in section \ref{FLtrustEval}. We denote by $c, \mu_k$ and $B$ the fraction of chosen agents at each round of algorithm, the learning rate, and the batch size respectively. At each round of \textit{trusted FedAvg}, the server first select $n = cN$ number of agents randomly. This is enforced by the fact that due to the large number of agents and the intermittent and bandwidth-constrained communications between the agents and the central server particularly in the case of wireless communications, it is not practical for all agents to participate in the model update at all rounds. The selected agents then can download the current global model and start the local training process denoted by the \textit{ModelUpdate} procedure. Each agent $i$ transmits a message to the server, denoted by $\mathbf{m}_i^{(k)}$. For a benign agent, $\mathbf{m}_i^{(k)} = \mathbf{w}_i^{(k)}$, while an adversary sends a message different from the update computed by running SGD on its local data. Finally, the server aggregates the local model updates received from the selected agents according to their trustworthiness and their dataset size reflected as the coefficients in at step 7. 
\begin{algorithm}
\caption{\textit{Trusted FedAvg}}
\label{trustedfedavg}
 \begin{algorithmic}[1]
 \renewcommand{\algorithmicrequire}{\textbf{Input:}}
 \renewcommand{\algorithmicensure}{\textbf{Output:}}
 \REQUIRE $n=c.N$, $\{q_i, i\in \mathcal{N}\}$, $\mu_k$, $B$
 \\ 
 \STATE Initialize $\mathbf{w}_{i}^{(0)}$
  \FOR{each round $k=1,2,...$}
  \STATE Server selects a subset $S_k$ of $n$ agents at random (agent $i$ is chosen with probability $q_i$)
  \STATE Server transmits $\mathbf{w}^{(k)}$ to all (chosen) agents
  \STATE Each agent $i\in S_k$ computes $\mathbf{w}_{i}^{(k+1)} = ModelUpdate(\mathbf{w}^{(k)}, \mu_k)$ and sends a message $\mathbf{m}_{i}^{(k+1)}$ to the server
  \STATE Server updates the trust values for each agent $i$:\\ $t_{Si}^{(k+1)} \leftarrow{} ComputeTrust(i, \{\mathbf{m}_{i}^{(k+1)}, i\in \mathcal{N}\})$
  \STATE Server aggregates the local updates:\\ $\mathbf{w}^{(k+1)} \leftarrow{} \sum_{i\in S_k} \frac{D_i t_{Si}^{(k+1)}}{\sum_{i\in S_t}D_i t_{Si}^{(k+1)}} \mathbf{m}_{i}^{(k+1)}$
  \ENDFOR\\
  $ComputeTrust(i, \{\mathbf{m}_{i}^{(k)}, i\in \mathcal{N}\})$:\\
  \STATE $\quad$ Server computes $I_{i}^{(k)}$ from \eqref{ghel1} or \eqref{ghel2}\\
  \STATE $\quad$ Server computes $t_{Si}^{(k)}$ according to \eqref{expected}\\
  $ModelUpdate(\mathbf{w}^{(k)}, \mu_k)$:
  \STATE $\quad$Initialize $\mathbf{\Psi}_{i,k} \leftarrow{} \mathbf{w}^{(k)}$ 
  \STATE $\quad \mathcal{B}$ = mini-batch of size $B$
  \STATE $\quad$\textbf{for} $b\in \mathcal{B}$ \textbf{do}
  \STATE $\qquad \mathbf{\Psi}_{i,k} \leftarrow{} \mathbf{\Psi}_{i,k} - \mu_k \nabla f(\mathbf{\Psi}_{i,k})$
  \STATE $\quad \textbf{end for}$
  \STATE $\quad$Return $\mathbf{w}_{i}^{(k)} \leftarrow{} \mathbf{\Psi}_{i,k}$
 \end{algorithmic}
\end{algorithm}

\subsection{Decentralized FL}
The main drawbacks of centralized FL are the scalability, connectivity  and single-point-of-failure issues. Moreover, next generation networks are expected to be enhanced by new forms of decentralized and infrastructureless communication schemes and device-to-device (D2D) multi-hop connections such as in UAV-aided networks \cite{gholami2020joint}. Within this scope, novel approaches are required to address decentralized (serverless) FL. While a number of research activities have focused on distributed learning algorithms \cite{shamir2014communication}, due to the special features of an FL setup in which the data is generated locally and remains decentralized and because of communication efficiency considerations, many existing approaches developed for distributed learning are not applicable to decentralized FL. 

In \cite{consensus-FL}, the centralized FL approach is extended for massively dense IoT networks that do not rely on a central server. Fig. \ref{FL}(b) shows the architecture of a decentralized FL setting. In \cite{consensus-FL}, agents perform training steps on their local data using SGD and consensus-based methods. At each consensus step, agents transmit their local model update to their one-hop neighbors. Each agent fuses the received messages from its neighbors and then feeds the result to SGD. We propose an attack-tolerant consensus-based FL algorithm by incorporating the trust concept into the consensus step. The proposed approach is given in Algorithm \ref{consensustrust}. In particular, let $t_{ij}^{(k)}$ denote the trustworthiness of agent $j\in\mathcal{N}_i$ evaluated at $i$. Similar to the centralized trusted FL approach, in step 7, agent $i$ aggregates the received updates from its neighbors with the weights of $\frac{D_j t_{ij}^{(k)}}{\sum_{j \in \mathcal{N}_i} D_j t_{ij}^{(k)}}$, i.e. the neighbors of $i$ with higher trust values contribute more to the aggregated model at $i$. Then, each agent updates its model independently using SGD on its local data.
\begin{algorithm}
\caption{\textit{Trusted Decentralized FL}}
\label{consensustrust}
 \begin{algorithmic}[1]
 \renewcommand{\algorithmicrequire}{\textbf{Input:}}
 \renewcommand{\algorithmicensure}{\textbf{Output:}}
 \REQUIRE $\mathcal{N}_i$, $\mu_k$, $\epsilon_k$
 \\ 
 \STATE Initialize $\mathbf{w}_{i}^{(0)}$ 
  \FOR{each round $k=1,2,...$}
  \STATE Agent $i$ receives the messages $\{\mathbf{m}_{j}^{(k)}\}_{j\in\mathcal{N}_i}$
  \STATE Agent $i$ updates the trust values for all its neighbors: \\
  $t_{i,j}^{(k)} \leftarrow ComputeTrust(i,j, \{\mathbf{m}_{j}^{(k)},j\in \mathcal{N}_i\})$
  \STATE $\mathbf{\Psi}_{i,k} \leftarrow{} \mathbf{m}_{i}^{(k)}$
  \FOR{each agents $j \in \mathcal{N}_i$}
  \STATE $\mathbf{\Psi}_{i,k} \leftarrow{} \mathbf{\Psi}_{i,k} + \epsilon_k \frac{D_j t_{i,j}^{(k)}}{\sum_{j \in \mathcal{N}_i} D_j t_{i,j}^{(k)}} (\mathbf{m}_{j}^{(k)}-\mathbf{w}_{i}^{(k)})$
  \ENDFOR
  \STATE Each agent $i$ computes:\\ $\mathbf{w}_{i}^{(k+1)} \leftarrow{} ModelUpdate(\mathbf{\Psi}_{i,k}, \mu_k)$ and sends $\mathbf{m}_{i}^{(k+1)}$ to all its neighbors
  \ENDFOR\\
  $ComputeTrust(i,j, \{\mathbf{m}_{i}^{(k)}, i\in \mathcal{N}\})$:\\
  \STATE $\quad$ Agent $i$ computes $I_{ij}^{(k)}$ from \eqref{ghel1} or \eqref{ghel2} \\
  \STATE $\quad$ Agent $i$ computes its local trust for $j$ ($\tau_{ij}^{(k)}$) from \eqref{expected}\\
  \STATE $\quad$ Agent $i$ computes its global trust for $j$ ($t_{ij}^{(k)}$) from \eqref{global}
  $ModelUpdate(\mathbf{w}_k, \mu_k)$:
  \STATE $\quad$ Initialize $\mathbf{\Psi}_{i,k} \leftarrow{} \mathbf{w}^{(k)}$ 
  \STATE $\quad$ $\mathcal{B}$ = mini-batch of size $B$
  \STATE $\quad$ \textbf{for} $b\in \mathcal{B}$ \textbf{do}
  \STATE $\qquad$ $ \mathbf{\Psi}_{i,k} \leftarrow{} \mathbf{\Psi}_{i,k} - \mu_k \nabla f(\mathbf{\Psi}_{i,k})$
  \STATE $\quad$ $ \textbf{end for}$
  \STATE $\quad$ Return $\mathbf{w}_{i,k} \leftarrow{} \mathbf{\Psi}_{i,k}$\\
 \end{algorithmic}
\end{algorithm}

\subsection{Trust Evaluation Method}
\label{FLtrustEval}
In this section we elaborate on the methods that are used in evaluating the trustworthiness of the network agents. These methods are embedded into the trust evaluation scheme used in the trust model in section \ref{sec:trustAgg}.
Throughout the FL protocols, in each iteration, the model parameters or their updates, corresponding to the local models of the agents are communicated within the network. These weights play an important role in determining the trust level of the agents. We enumerate multiple methods that assign trust values to the agents based on the communicated weight updates: 

\begin{itemize}
    \item \textit{Clustering-based Method}: In this method, the trustor entity $i$ compares the messages it has received from trustee $j$, to all the other messages it has received from the other parties. Formally, for each neighbor $j$, party $i$ computes: 
    
    \begin{equation}
        \operatorname{dev}_{i j}^{(k)}=\sum_{l \in \mathcal{N}_{i}^{+}} \frac{\left|\left|w_{l}^{(k)}-w_{j}^{(k)}\right|\right|_2^{2}}{\left|\mathcal{N}_{i}^{+}\right|}\label{dev}
    \end{equation}
    
    where ${\left|\mathcal{N}_{i}^{+}\right|}$ is the set of the neighbors of agent $i$ containing itself. Then, for each trustee $j$, it will benchmark the value of  ${dev }_{ij}^{(k)}$ against a multiple of the median of all the deviations:
    
    \begin{equation}
    I_{i j}^{(k)}=\left\{\begin{array}{ll}
    1 &  {dev }_{ij}^{(k)} \leq {th}_{i}*\text { median }\left(\left\{\operatorname{dev}_{i j}^{(k)}\right\}\right) \\
    0 & {o.w.}
    \end{array}\right. \label{ghel1}
    \end{equation}
    
    This way, by adjusting the value of $th_i$ at iteration $k$, the trustor party can decide not to trust those parties from which it has received too-far-away messages. 
    
    \item \textit{Distance-based Method}: In this method, the trustor party $i$ computes the distance between its local model and the model at trustee $j$, and uses this distance as a measure to assign trust values to its neighbors. Formally, party $i$ computes the distance between the message it has received from party $j$ in the previous and the current iterations; i.e.
    
    \begin{equation}
        d_{i j}\left(w_{i}^{(k-1)}, m_{j}^{(k-1)}\right)=\left\|w_{i}^{(k-1)}-m_{j}^{(k-1)}\right\|_{2}
    \end{equation}
    
        and, 
        
    \begin{equation}
        d_{i j}\left(w_{i}^{(k-1)}, m_{j}^{(k)}\right)=\left\|w_{i}^{(k-1)}-m_{j}^{(k)}\right\|_{2}
    \end{equation}

Then party $i$ computes the difference between the two computed distances and decides on the value of $I_{ij}^{(k)}$ as follows: 

\begin{equation}
    I_{ij}^{(k)} = \mathbb{I}_{\left\{ d_{i j}\left(w_{i}^{(k-1)}, m_{j}^{(k-1)}\right) - d_{i j}\left(w_{i}^{(k-1)}, m_{j}^{(k)}\right) \geq 0\right\}} \label{ghel2}
\end{equation}

In other words, if node $j$ is has a benign behavior, then the in one iteration of the protocol, its local model must have shifted towards the local model of party $i$. If this is not the case then party $j$ has to be malicious or must be communicating incorrect message to $i$.
    
\end{itemize}

\subsection{Numerical Results}

\begin{figure}
\begin{center}
\begin{minipage}[h]{0.45\textwidth}
\includegraphics[width=1\textwidth]{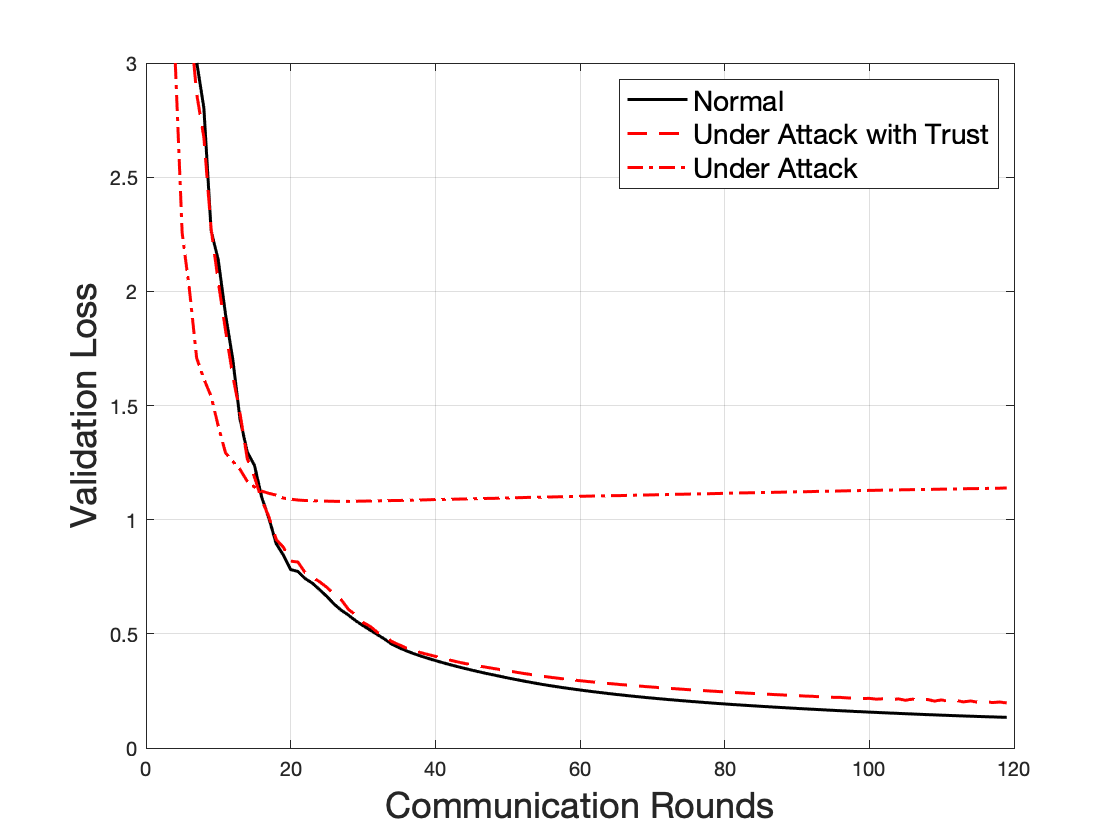}
  \caption{Effect of trust on resilience against attacks}
  \label{trust_fl_dec}
\end{minipage}
\end{center}
\end{figure}

For the simulation setup, we adopt the settings of \cite{consensus-FL}. Our implementation of the FL process and the validation dataset are both based on \cite{WinNT}.  We implement an attack at the presence of $80$ nodes participating in the FL task with $10 \%$ of the nodes being corrupt. The attacker parties will generate the poisoned model by choosing the weights randomly in the range $(a*w_{min}, a*w_{max})$ where $w_{min}$ and $w_{max}$ are the minimum and maximum of the weights they receive from their neighbors. We implement this attack under two \textit{mild} and \textit{hard} settings corresponding to the case where $a=1$, and $a=2$ respectively. We have depicted $120$ epochs of the process when the attack is mild. Fig. \ref{trust_fl_dec} shows how incorporating trust into the decentralized federated learning framework can protect the protocol from being invaded by malicious parties. In the absence of the trust mechanism, under corrupt network agents, the validation loss will not converge to the correct value corresponding to when the nodes are operating normally. Therefore, the performance of the trained model on test data degrades significantly, even when implementing the mild version of the attack. However, when the trust model is in place the trained model will converge to that of the normal implementation. We note that the validation loss after $120$ epoch converges to $0.14$, and $0.18$ for the normal (without any attacks) and trust-aware models, and diverges from $1.2$ for the attacked unprotected model. The $0.14$ validation loss corresponds to a $90\%$ of accuracy on the test data.

\section{Case Study II: Trust-Aware Network Service Placement}
\label{sec:SCE}
Software-Defined Networking (SDN) and Network Function Virtualization (NFV), are two complementary technologies that together have revolutionized the the field of communications and computer networks. SDN, decouples the control logic of the of the network from its forwarding layer, dividing the network into two separate systems with different dynamics; i.e. control plane, and data plane. The data plane only consists of servers that are responsible for forwarding the data packets and are controlled and programmed by the SDN controllers that reside within the control plane. The SDN controllers in the control plane maintain a global view of the network devices and are in charge of making all the decisions such as routing, resource allocation, security preserving, etc. in the network in a logically-centralized manner. The servers of the data plane are connected to the SDN controllers with a separate protocol (e.g. OpenFlow), through which they receive the operating instructions and also report back their latest status. NFV allows for decoupling the network functions from the dedicated hardware and allows for realizing such function through software. These virtualized network functions (VNFs) can be implemented on commodity hardware. This not only reduces the cost of network operation but also allows for allocating the required infrastructure resources on demand at scale. Each complex network service is a set of network functions (e.g. load balancing, firewall, intrusion detection, etc. ) that are stitched together with some logical links that determine the order in which they need to process the network packets. Such a representation is denoted as VNF forwarding graph (VNF-FG). The network functions will be run on the infrastructure servers (consuming processing power), and the traffic flowing between the VNFs will be mapped onto the infrastructure paths (consuming bandwidth); i.e the infrastructure is in fact a large set of physical servers and communication links between them with a pool of resources (processing, bandwidth, etc.). This problem is known as the VNF-FG placement, network service placement, or service function chain (SFC) embedding \cite{9289885} \cite{8768602}. The optimal solution to this problem is one that allocates the resources to the network services in the most cost-efficient manner, while satisfying the design requirements input as constraints to the model.

\begin{figure}
\begin{center}
\begin{minipage}[h]{0.42\textwidth}
\includegraphics[width=1\textwidth]{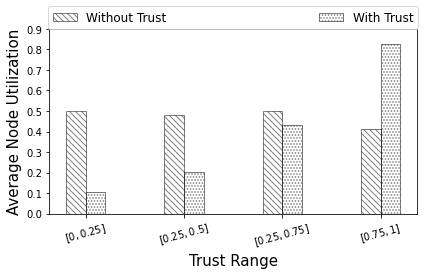}
  \caption{Server Utilization Rate versus Trust Level}
  \label{trust_SCE_sol}
\end{minipage}
\end{center}
\end{figure}

In \cite{8768602}, and \cite{9289885}, we addressed the above problem while considering security-oriented constraints. We used trust as a metric to capture the required security level of the network functions and the capability of the infrastructure servers in providing the corresponding security level. This very well matches with the centralized model presented in section \ref{sec:trustAgg}. Each of the network servers play the role of the collaborating agents that communicate with the central entity i.e. SDN controller (or the orchestrator) which is in charge of calculating the trust estimates based on the evidence it has obtained, and updating the servers' trust values. The existence of a protocol for communications between the data plane and the control plane makes it easy for the SDN controller to access the telemetry data and status of the physical servers and yields the provision of evidence realizable. Fig. \ref{trust_SCE_sol} shows how the trust-aware solution correctly allocates more load to more trusted servers when comparing with the plain model. We do not include the rest of results here due to the space limitation. For a detailed analysis of the role of trust in SFC embedding the interested reader is referred to \cite{8768602}, and \cite{9289885}.
 \begin{figure}
\begin{center}
\begin{minipage}[h]{0.4\textwidth}
\includegraphics[width=1\textwidth]{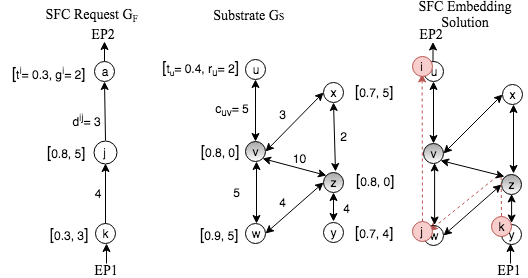}
  \caption{Example of trust-aware network service placement}
  \label{trust_SCE}
\end{minipage}
\end{center}
\end{figure}
\section{Challenges}
\label{sec:problem}

Several challenges arise in practice, when designing a realistic trust aggregation framework. In this section we enumerate the existing obstacles on the way of incorporating trust into the decision making process for the next-generation networked systems and then propose our ideas for tackling these challenges.

\subsection{Private Communication-efficient Trust Aggregation}
As explained in section \ref{sec:intro}, in the next-generation NetCPS, the information is crowd-sourced by the CPS devices. Further, we justified the importance of the existence of a framework for inferring and aggregating the trustworthiness of the networked agents, and described two such \textit{centralized} and \textit{decentralized} structures. However, there are two major drawbacks from these schemes concerning the communication-efficiency and privacy of them. In the centralized scheme, the agents have to share their data with the central entity who is in charge of making the trust assignment decisions. However, on one hand, as the number of CPS devices increases and the data grow in size it will soon become bandwidth-inefficient to transmit the data to the central entity (especially, in proactive settings). On the other hand, it very well may be the case that the data gathered by or generated at a CPS agent contain sensitive or private information that the agent is not willing to share with another party. Especially, in wireless networks where it might take several hops for each agent to reach the central server; This not only increases the chance of privacy violation, but also may increase the delay of communication between the agent and the central server. With respect to the decentralized architecture, these issues are not resolved completely. Under these circumstances, the parties will have to communicate with one another and exchange their views regrading the trustworthiness of the fellow peers, rather than directly communicating with a centralized server. Although this may have the advantage of reducing the communication delays and lead to partial bandwidth efficiency, but may increase the net volume of information communication required to reach the same level of accuracy as in the centralized scheme. Moreover, the fact that the decentralized approach will have the devices share their opinions on the trustworthiness of other fellows, may make the privacy issue even worse. Additionally, as discussed in section \ref{sec:intro}, due to potential competition between the networked agents, the parties might not be willing at all to respond to any queries by the fellow peers. 
\subsection{Trust Update Freshness}
Trust is a metric that may evolve dynamically. Therefore, the resolution of updating the trust estimates in the network has to be adjusted carefully and tailored to the frequency of the trust-based decision makings; i.e. the most up-to-date trust estimates need to be available whenever required by the corresponding entities. In other words, it is of paramount importance that the trust information at the evaluating entities is fresh and an indicator of the current state of the network. We note that this argument is different from minimizing the latency of the trust estimates in the network and captures a separate requirement. \textit{Age of information (AoI)}, is a recently-introduced metric that aims at maximizing the freshness of data in multi-server systems, and has received a lot of attention in academia, due to its effectiveness in guaranteeing the freshness of status updates. We believe the path to an effective solution to the last challenge, crosses the AoI metric. 

\subsection{Quantifiable Trust}
Another bottleneck in designing trust-aware mechanisms is the restrictiveness of quantifiable trust. Although there exists a large number of works in the literature concerning the notion of trust with various approaches, most of them maintain the qualitative perspective, so there are few works that study the quantification of trust. Among those few, most of them take the quantitative values for trust as granted without providing a detailed analysis on how to obtain such values. We believe this shortcoming stems from the abstractness of the notion of trust, the complexity of trust evaluation, and its application-specificity. Trust may be a composite of several metrics such as, reliability, availability, resilience, adaptability, reputation etc., and taking into account all these factors for all functions performed within the network, will add to the complexity of trust evaluation. Moreover, it is impossible to define trustworthiness against all types of intrusions and security threats. Therefore, trust has to be viewed at the system level rather than a specific low level preventive measure against specific attacks. In \cite{8329991}, the author has interestingly enumerated the studies on trust quantification in different fields and also has provided a novel method for evaluating trust based on quantitatively modeling and measuring ability,  benevolence, and integrity, as its indicators. 
\section{Conclusions}
\label{sec:conclusions}

In this paper we studied the importance of trust in various types of decision makings in the next-generation cyber-physical systems. We described the terminology of trust and its types and presented a centralized and a decentralized structure for trust aggregation. Next, we invoked several interesting use-cases were trust could significantly boost the quality of the decisions made in the networked systems resulting in higher security and resiliency of the discussed methods. Specifically, we examined the centralized and the decentralized federated learning frameworks and showed how incorporating trust into agent participation decisions can improve the resiliency of the framework against various types of attacks. We further discussed how a trust-aware NFV resource allocation scheme may increase the security measures of a cloud network provisioning player. 
We then explained multiple challenges that arise against incorporating trust within the decision making  framework in practice. Dealing with these challenges are among the future directions of our research in this topics. Specifically, we aim to come up with a framework that ensures the timeliness of trust estimates using the AoI concept. Moreover we plan to model the trust aggregation problem as a learning task and  use the capacities of Federated Learning to ensure a privacy-preserving communication-efficient trust inference method. 




%
\bibliographystyle{IEEEtran}
\bibliography{ref}

\end{document}